\documentstyle[11pt,paspconf]{article}

\def \kms{\ifmmode\,{\rm km}\,{\rm s}^{-1}\else km$\,$s$^{-1}$\fi} 
\def\halpha{H$\alpha$}
\def \ifm#1{\relax\ifmmode#1\else$\mathsurround=0pt #1$\fi} 
\def \hmpc{\,h\ifm{^{-1}}{\rm Mpc}} 
\def \etal{{\sl et al.}} 
 
\def \eg{{\it e.g.}} 
\def \arcmin{\hbox{$^\prime$}}

\def \gtsima{$\, \buildrel > \over \sim \,$} 
\def \ltsima{$\, \buildrel < \over \sim \,$} 
\def \simgt{\lower.5ex\hbox{\gtsima}} 
\def \simlt{\lower.5ex\hbox{\ltsima}}

\def \bfV{{\bf V}}

\def \nhat{\ifmmode {\hat{\bf n}}\else${\hat {\bf n}}$\fi}
\def \kmsmpc {\kms\ {{\rm Mpc}}^{-1}}

\def \littleprime{\ifmmode{\scriptscriptstyle \prime } 
     \else{\hbox{$\scriptscriptstyle \prime$ }}\fi} 
\def \littless{\ifmmode{\scriptscriptstyle s } 
     \else{\hbox{$\scriptscriptstyle s $ }}\fi} 
\def \littlemm{\ifmmode{\scriptscriptstyle m } 
     \else{\hbox{$\scriptscriptstyle m $ }}\fi} 
\def \littlecirc{\ifmmode{\scriptscriptstyle \circ } 
     \else{\hbox{$\scriptscriptstyle \circ $ }}\fi} 
\def \littlehour{\ifmmode{\scriptscriptstyle h } 
     \else{\hbox{$\scriptscriptstyle m $}}\fi}
\def \toph{\raise .9ex \hbox{\littlehour}} 
\def \hourpoint{\hbox to 2pt{}\rlap{\hskip -.5ex \toph}.\hbox to 2pt{}} 
\def \topemm{\raise .9ex \hbox{\littlemm}} 
\def \magpoint{\hbox to 2pt{}\rlap{\hskip -.5ex \topemm}.\hbox to 2pt{}} 
\def \arcss{\raise .9ex \hbox{\littless}} 
\def \arcsspoint{\hbox to 1pt{}\rlap{\arcss}.\hbox to 2pt{}} 
\def \sdeg{\raise .9ex \hbox{\littlecirc}} 
\def \degpoint{\hbox to 1pt{}\rlap{\sdeg}.\hbox to 2pt{}} 
\def \arcmin{\raise .9ex \hbox{\littleprime}} 
\def \arcminpoint{\hbox to 1pt{}\rlap{\arcmin}.\hbox to 2pt{}} 
\def \arcsec{\raise .9ex \hbox{\littleprime\hskip-3pt\littleprime}} 
\def \arcsecpoint{\hbox to 1pt{}\rlap{\arcsec}.\hbox to 2pt{}} 
\def \plotfiddle#1#2#3#4#5#6#7{\centering \leavevmode 
\vbox to#2{\rule{0pt}{#2}} 
\includegraphics{#1}} 
 
\begin{document}

\title{First Results from the Shellflow Survey}

\author{ 
St\'ephane Courteau$^{1}$, Jeffrey A. Willick$^{2}$, Michael A. Strauss$^{3}$,  
David Schlegel$^{3}$, Marc Postman$^{4}$ 
} 
 
\affil{(1) \ Herzberg Institute of Astrophysics, Victoria, BC}
 
\affil{(2) \ Stanford University, Department of Physics, Stanford, CA} 
 
\affil{(3) \ Princeton University Observatory, 
 Princeton, NJ 08544} 
 
\affil{(4) \ Space Telescope Science Institute, Baltimore, MD} 
 


\begin{abstract}
We present preliminary results from the Shellflow program, an all-sky 
Tully-Fisher (TF) survey of 297 Sb$-$Sc with redshifts
between 4000 and 7500 \kms.
The program was designed to ensure uniformity of the data between
observing runs and between telescopes, thereby eliminating the
possibility of a spurious bulk flow caused by data inhomogeneity.
A simple bulk flow model suggests that the 
Shellflow galaxies are nearly at rest 
with respect to the cosmic microwave background (CMB). 
Taken at face value, this result suggests that most of the $\sim 600$
\kms\ motion of the Local Group with respect to the CMB is due to 
material within $\sim 60\hmpc$ --- a striking confirmation of the 
gravitational 
instability picture and the notion of homogeneity in the Universe.
Differences between our Shellflow ``null'' result and claims of larger 
bulk motions based on analyses of the MarkIII catalog are 
possibly due to small inhomogeneities between the sub-samples in MarkIII.
Our preliminary 
Shellflow bulk velocity, $V_{bulk} = 80 \pm 150$ \kms\ 
in the CMB frame, must still be refined with Monte-Carlo simulations and 
tidal field analysis.  

\end{abstract}




\section{Introduction}

It is of great cosmological importance to identify the volume of space,  
centered on the Local Group, which is at rest with respect to the CMB.
Very large scale fluctuations are required to move ever larger volumes 
of space in the standard gravitational instability model.  
In standard Cold Dark Matter cosmologies, 
the volume of space bounded by the nearest superclusters (Great 
Attractor, Pisces-Perseus, Coma) is expected to define an inertial frame 
nearly at rest with respect to the CMB.  The distribution of the matter within 
this volume should explain the $\sim 600$ \kms\ motion of the Local Group in 
the CMB frame.  The detection of a large amplitude 
($V_{bulk} \simgt 700$ \kms) 
on scales exceeding 8000 \kms\ by Lauer \& Postman (1994), followed by 
recent measurements with similar amplitude by Willick (these 
proceedings) 
and the SMAC team (Smith, these proceedings), have not only challenged the 
notion that the bulk flow on large scales is small, but have also pushed 
cosmological models to the breaking point.  

Moreover, 
there are contradictory claims for the observed  
bulk flow within a sphere of 6000 \kms, 
and whether it is generated by internal or external mass 
fluctuations.  
The most recent POTENT reconstruction of the MarkIII velocities 
(Dekel \etal\ 1999) 
found a bulk velocity within 6000 \kms\ of
$370 \pm 125 \kms$ in the  
CMB frame towards Supergalactic $(L,B)=(165^\circ,\, -10^\circ)$. Dekel
\etal\ (1999) argue that this motion
is generated by the {\it external} mass distribution on very large
scales (see also Courteau \etal\ 1993).  
However, other investigations using nearly homogeneous samples of galaxies 
within and beyond $\sim 60\hmpc$ find motion consistent with the amplitude  
and direction of the CMB dipole (Giovanelli, these proceedings).  New local 
peculiar velocity measurements based on the brightness of supernovae Type Ia
(Riess, these proceedings) and surface brightness fluctuations (Tonry and Dressler,
these proceedings) are also consistent with little bulk motion at $60\hmpc$. This 
suggests that the reflex motion of the Local Group could be explained by 
material contained {\it within} $60\hmpc$. 

The controversy over the observed bulk flow within $60\hmpc$ stems, 
in large part, from 
our inability to combine the various galaxy distance samples used in flow 
studies into a single homogeneous catalog.
None of the surveys within $60\hmpc$ had sampled the {\it entire} sky 
uniformly\footnote{Earlier attempts include David Schlegel and Josh Roth's 
 thesis work.}.
In an effort to overcome this situation, we devised a new study that 
would take advantage of Northern and 
Southern NOAO facilities to map the motions of a shell of galaxies with 
significant overlap between the hemispheres, and with existing 
peculiar velocity surveys.
Our new {``\it Shellflow''}
survey was designed to provide {\it precise\/} and {\it uniform\/} 
photometric and spectroscopic data over the whole sky,
and thus to remove the uncertainties associated with matching 
heterogeneous data sets. 
With Shellflow, and further calibration of existing TF data  
sets, we expect a high-accuracy detection of the bulk flow amplitude  
and a better constrained characterization of the tidal field at 6000 \kms.

\section{Observational Strategy}
 
The Shellflow sample is drawn from the Optical Redshift Survey sample
of Santiago
\etal\ (1995; ORS). The ORS sample consists of all galaxies
in the UGC, ESO, and ESGC Catalogs  with
$m_{\rm{B}} \leq 14.5$ and $|b| \geq 20^\circ.$ 
We initially selected all non-interacting Sb and Sc galaxies
in the ORS with 
redshifts\footnote{We actually define three subsamples complete in that range 
with different definitions of redshift: measured in the Local Group frame, 
the CMB frame, and after correction for peculiar velocities according to 
the {\sl IRAS\/} model of Yahil \etal\ (1991) with $\beta = 1$; if we chose 
only one of them, the sample would decrease in size by 20\%.} between  
4000 and 7500 \kms,  inclinations between  
$45^\circ$ and $78^\circ$, and with Burstein-Heiles (1982) 
extinctions $A_B\leq$  
0\magpoint30. This produced a catalog of more than 300 
objects.  All galaxies were then visually inspected on the Digitized POSS
plates; those with bright foreground stars and obvious disturbances were 
excluded, yielding a final sample of 297 Shellflow galaxies.
No pruning was done of galaxies not matching idealized 
morphologies beyond the restriction on Hubble type and inclination. 

The data were collected between March 1996 and March 1998 at NOAO,
using V and I-band photometry and \halpha\ rotation curves. 
Data taking and reduction techniques follow the basic guidelines of 
previous optical TF surveys (\eg\ Courteau 1996, 1997; Schlegel 1995).
The V and I-band imaging was obtained at the CTIO and KPNO 0.9m 
telescopes.  The photometric calibration is based on the Kron-Cousins 
system; data taken on nights with standard star photometric scatter 
greater than 0\magpoint020 were excluded.  The Kron-Cousins 
system also allows direct matching with two largest I-band TF 
samples to date (Mathewson \etal\ 1992, Giovanelli \etal\ 1998). 
The \halpha\ spectroscopy was obtained mostly in photometric 
conditions 
with the RC spectrographs at the CTIO and KPNO 4m telescopes. 
Typical integrations were $\sim 900$s and $\sim 1800$s for imaging 
and spectroscopy respectively.  

Forty-one galaxies were imaged at both CTIO and KPNO, and we have
repeat imaging from a given telescope for 106 galaxies.  In addition,
we observed 27 galaxies spectroscopically from both CTIO and KPNO, and
obtained duplicate spectra from a given telescope for 38 galaxies.
The total magnitudes and rotational line widths reproduce to within
0\magpoint06 and 3 \kms\ (rms deviations) respectively, with no
systematic effects seen between hemispheres or between runs.
Moreover, all data reduction was done independently by Courteau and
Willick using different software and methodology; the results between
the two agree to within the errors quoted above.  The high level of
accuracy in our Shellflow data meets our requirement for a measurement
of a significant bulk flow result.



\begin{figure} 
 \plotfiddle{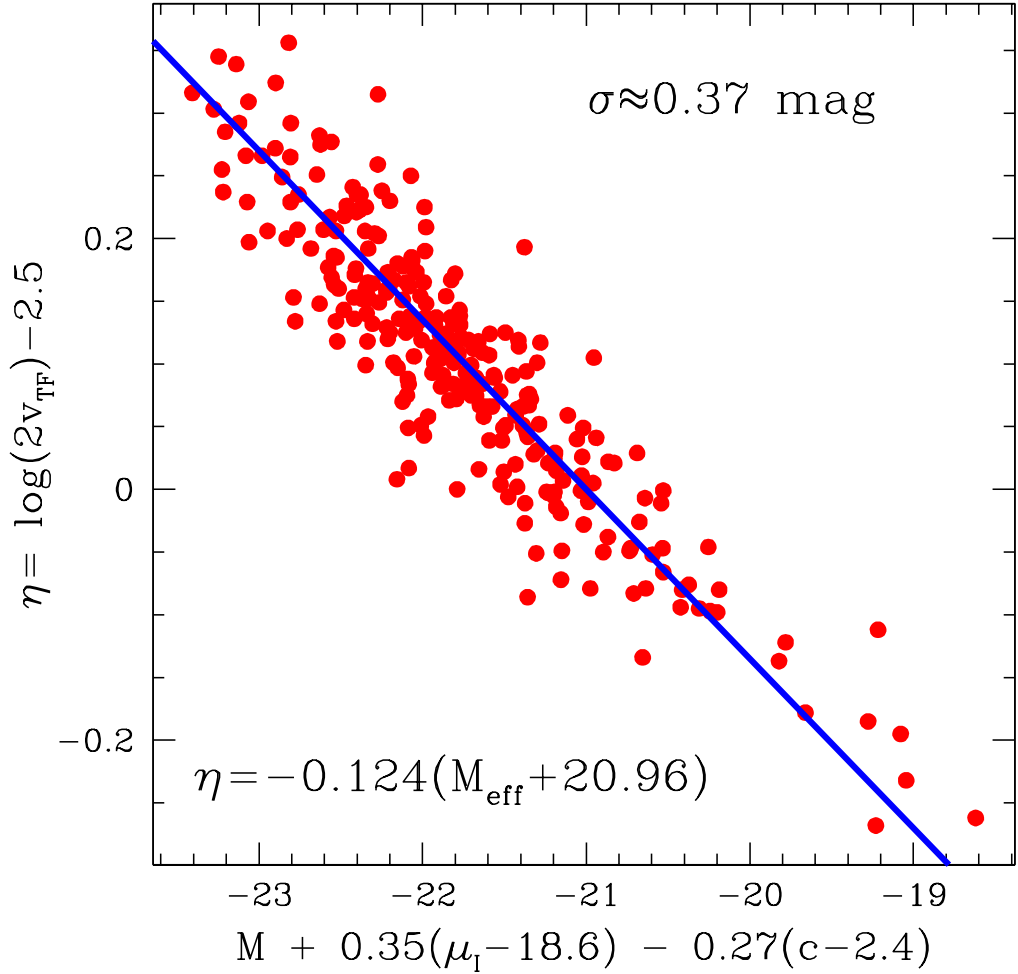}{6cm}{0}{70}{70}{-230}{-240} 
 \vspace{6cm} 
 \caption{Tully-Fisher calibration of the Shellflow sample.
 This figure includes all 297 Shellflow galaxies.  The slope
 and scatter are comparable to current modern I-band TF 
 investigations.  The absolute magnitudes are corrected
 for the small surface brightness 
 and concentration index dependences of the TF relation. 
 As mentioned in the text, the reality of these dependences
 is still under investigation by ourselves and other workers.
 The results of our flow
 fits are, however, independent of whether or not these additional
 correlations are included.}
\end{figure}

\section{Analysis and Results} 

Following the approach of Lauer \& Postman (1994), we use the sample 
itself to  
calibrate the distance indicator relation; this mitigates the need to  
tie the sample to external TF calibrators such as clusters (although 
it precludes measurement of a monopole term in the velocity field). 
We choose the ``inverse'' form of the TF relation, which makes Malmquist 
and selection bias effects negligible (cf., Strauss \& Willick 1995). 
A maximum likelihood analysis for a pure Hubble expansion model (no bulk
flow) allows us to constrain $\sigma^{int}$, the intrinsic TF scatter,
and $\delta v$, the typical error in measuring raw rotation velocities. 
We hardwire $\sigma_{phot}$, the total error associated with photometry 
(including inclination errors), to 0.15 mag, which is a small
fraction of the error budget.  Two other parameters constrain
the dependence of the TF relation on surface brightness and shape of
the rotation curve.  The best TF fit is achieved for $V_{rot}$ measured near
1.7 disk scale lengths, similar to but somewhat smaller
than the values found by Courteau (1997) and Willick (1999).
A small surface brightness dependence is also found
for the optimized TF relation, although this could be induced by 
the particular choice of luminosity profile fitting that we have
adopted (see also the discussion by Willick [LP10K], these proceedings;
Courteau \etal\ 2000).
 
\vfill\eject

For this analysis, we adopt a model of {\bf Hubble expansion plus a 
bulk flow\/} to compute absolute from apparent magnitudes:  

\[ d ({\rm Mpc}) = \frac{1}{H_0} \left(cz - \bfV_B\cdot\nhat\right) \,, \]  

\noindent 
where $H_0=100\,\kmsmpc$ and $cz$ is the measured redshift in \kms.  
We measure $cz$ in either the {\bf CMB\/} frame or the {\bf LG\/} frame.  
(A different choice of Hubble constant would shift
the zero point of the derived TF relation but would have
no effect on our bulk flow fits.)

The TF fit slope and scatter are comparable to recent 
I-band investigations (\eg\ Giovanelli \etal\ 1998).  
The Shellflow TF relation is shown in Fig.~1.  We find no 
dependence of the internal extinction correction on luminosity, 
and the TF parameters are virtually unchanged 
whether one adopts Burstein-Heiles (BH), or Schlegel, 
Finkbeiner, \& Davis 1998 (SFD) Galactic extinctions. 
The optimization with respect to all TF parameters and bulk flow
velocity components (V$_x$,V$_y$,V$_z$), yields the following fits: 

\begin{table}
\caption{\bf Best-fit Velocity Components (\kms)} 
\begin{center} 
\begin{tabular}{rrr|l}  
\multicolumn{4}{c}{{\bf SFD extinctions}} \\ \hline\hline  
\noalign{\vskip 3pt}
V$_z$ \  & V$_x$ \  & V$_y$ \  & Frame \\
\noalign{\vskip 3pt}
 \hline  
\noalign{\vskip 5pt}
$  47.5$&$ -37.7$&$  28.6$&CMB   \\  
$-257.1$&$ -98.9$&$ 566.1$&LG    \\  
\end{tabular}  

\vskip 10pt

\begin{tabular}{rrr|l}  
\multicolumn{4}{c}{{\bf BH extinctions}} \\ \hline\hline  
\noalign{\vskip 3pt}
V$_z$ \  & V$_x$ \ & V$_y$ \   & Frame \\ 
\noalign{\vskip 3pt}
\hline  
\noalign{\vskip 5pt}
$  58.8$&$ -38.0$&$  46.6$&CMB   \\   
$-242.6$&$ -99.2$&$ 584.5$&LG    \\   
\end{tabular}  
\end{center} 
\end{table}

\begin{figure}
 \plotfiddle{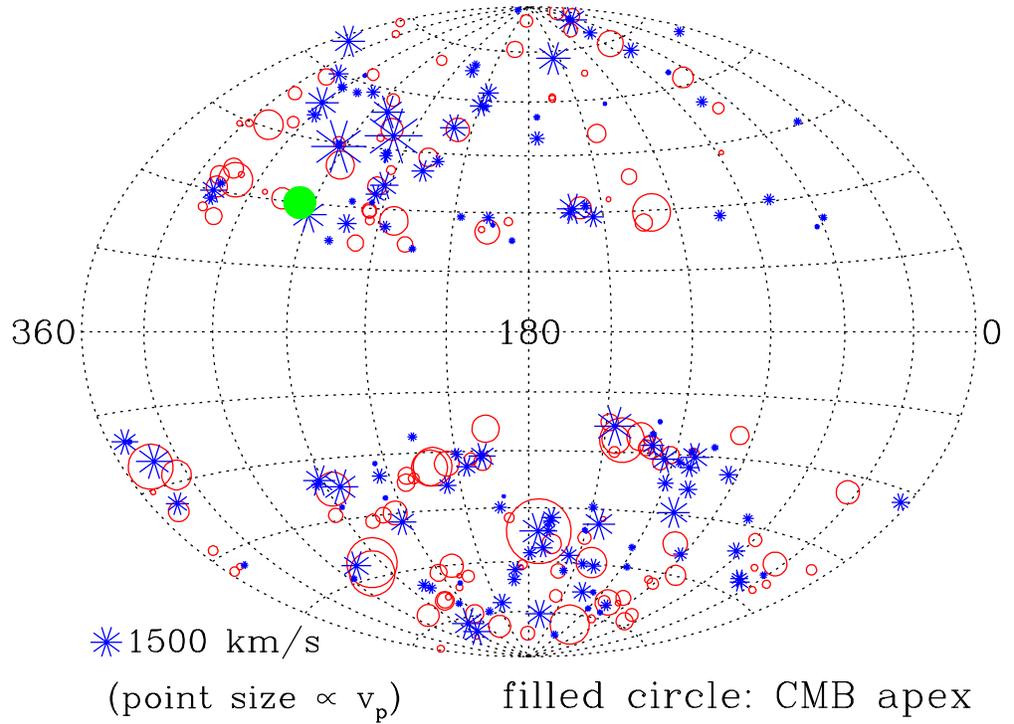}{6cm}{0}{70}{70}{-230}{-190} 
  \vspace{2cm} 
 \caption{Plot of peculiar velocities in the CMB frame.  
 The point size is proportional to the velocity amplitude,
 as indicated by the key at the lower left. 
 The circles and asterisks represent 
 inflowing and outflowing objects, respectively.  
 The two point types are well-mixed at all positions
 on the sky, indicating the absence of any significant
 bulk flow in the CMB frame. 
 }
\end{figure} 

The typical velocity errors, computed by holding the other 
parameters fixed, are $\Delta$V$_x = \pm 110$ \kms, 
$\Delta$V$_y = \pm 90$ \kms, 
$\Delta$V$_z = \pm 70$ \kms, 
for an upper limit on the total amplitude equal to
$\Delta$V $\leq 150$ \kms $(1\sigma)$.  Numerical simulations to
assess the full covariance matrix of the errors will be presented
elsewhere. 

Modelling the Shellflow galaxy peculiar velocities with  
a Hubble expansion and a bulk flow, one finds a low bulk motion
of $80 \pm 150$ \kms\ with respect to the CMB.
Correspondingly, the Shellflow volume moves at $\sim 600$ \kms\ in the 
frame of the LG\footnote{See Courteau \& van den Bergh (1999) for caveats
regarding transformations to the Local Group reference frame.}.
This corresponds to the reflex motion of the Local
Group motion with respect to the CMB. The directional components 
of the flow are poorly constrained since its amplitude is so low.

Our result is in agreement with other newly-presented results
at this workshop, derived from surface brightness fluctuation techniques 
(Tonry \& Dressler; V$_{bulk}=289\pm137$ \kms\ at 3000 \kms), nearby SNIa 
($V_{bulk}$ consistent with zero at 6000 \kms (Riess, these proceedings),
and previous TF analyses by Giovanelli and collaborators. 
Taken together these results suggest that most of the mass responsible 
for the motion of the LG lies within 6000 \kms.  Thanks to many new 
peculiar velocity surveys, the picture of cosmological 
bulk flows seems to imply ``convergence'' of the flow field within  
6000 \kms\ to the CMB dipole value. 
However, this simple picture is challenged by a mix of low and  
high amplitude bulk flow measurements on larger scales.  If the flow is nearly 
quiet at $\sim 6000-7000$ \kms, one is hard-pressed to explain the  
reported high amplitude flows on larger scales. These issues 
are discussed elsewhere in these proceedings.

\section{The MarkIV Catalog of Galaxy Peculiar Velocities}
 
Two of us (JW \& SC $+$ collaborators) 
have recently combined the major distance-redshift surveys from both  
hemispheres (published before 1994) into a catalog of 3100 galaxies 
(Willick \etal\ 1997, the ``MarkIII Catalog'').  
However, full homogenization at the $2-3$\% level, 
the minimum required for an accurate bulk flow detection at 6000 \kms,  
was not achieved in MarkIII due to irreducible uncertainties  
associated with matching disparate TF datasets.
Furthermore, a revised calibration 
of the MarkIII TF zero-points based on maximal  
agreement with the peculiar velocities predicted by the IRAS 1.2Jy redshift  
survey suggests a possible source of systematic error for the 
data sets which cover the PP cone (Davis, Nusser, \& Willick 1996, 
Willick \& Strauss 1998 [VELMOD]). 
This uncertainty has not seriously affected mass density 
reconstructions (Dekel \etal\ 1999) but it could lead to spurious 
estimates of the bulk flows on larger scales.  

MarkIII was originally tied to the North/South cluster calibration of 
Han \& Mould (1992).  We now plan a revised MarkIV calibration based on 
Shellflow.  Because of the overlap with existing surveys at comparable 
depth, Shellflow will be of fundamental importance in tying these datasets   
together in a uniform way. With this more accurate global calibration,
we expect that the MarkIV catalog will be better suited than
its predecessors for studying the velocity and density fields 
in the local universe.

\section{Summary}

\begin{itemize}
\item The Shellflow velocity components in the CMB frame are small. 
      {\it There is no significant bulk motion of the Shellflow sample
      relative to the CMB.}
\item The Shellflow velocity components are large when the fit is done 
      in the LG frame.  This represents the reflex of the LG motion 
      with respect to the CMB. 
\item Results are insensitive to BH versus SFD Galactic extinctions.
\item The Shellflow bulk motion reported here has not been fully tested 
      against Monte-Carlo simulations or using a complete tidal field
      analysis. A comprehensive study is under way. 
\end{itemize}

\acknowledgments

We wish to thank various students and postdocs who have contributed 
to some of the Shellflow reductions. In Victoria, Shelly Pinder and
Yong-John Sohn, and Josh Simon, Felicia Tam, and Marcos
Lopez-Caniego at Stanford. MAS acknowledges support from Research
Corporation and NSF grant AST96-16901. JAW acknowledges support
from Research Corporation and NSF grant AST96-17188. 




%

\end{document}